\documentclass[useAMS,usenatbib]{mn2e}

\usepackage{times}
\usepackage{graphicx,ulem}

\newcommand{\beq}{\begin{equation}}
\newcommand{\eeq}{\end{equation}}

\newcommand\Sigcr{\Sigma_{\rm cr}}

\newcommand\al{\alpha}

\def\arcminf{\!\!^{\prime}}

\def\apj{ApJ}
\def\mnras{MNRAS}
\def\apjl{ApJL}
\def\aj{AJ}
\def\aap{A\&A}
  
\def \araa{ARAA}
\def \aaps{AAPS}
\def\pasj{PASJ}

\def\gs{\mathrel{\lower0.6ex\hbox{$\buildrel {\textstyle >}\over{\scriptstyle \sim}$}}}
\def\ls{\mathrel{\lower0.6ex\hbox{$\buildrel {\textstyle <}\over{\scriptstyle \sim}$}}}

\begin{document} 
\title{The dynamical state of RX J1347.5-1145 from a combined strong lensing and X-ray analysis}
\author[Miranda et al.: RX J1347-1145 strong lensing and X-ray analysis]{M. Miranda$^1$\thanks{E-mail: solar@physik.unizh.ch}, M. Sereno$^1$, E. De Filippis$^{2,3,4}$, M. Paolillo$^{3,4}$\\
$^1$Institute for Theoretical Physics, University of Z\"urich,        
Winterthurerstrasse 190, CH-8057 Z\"urich, Switzerland \\    
$^2$INAF -Italian  National Institute of Astrophysics, via del Parco Mellini,
Rome, Italy \\
$^3$Dipartimento di Scienze Fisiche, Universit\`a di Napoli, 
               Via Cinthia, 80126 Napoli, Italy and\\ 
$^4$INFN-Napoli Unit, Via Cinthia, 80126 Napoli, Italy\\
}
\pubyear{2007}       
\maketitle 
\begin{abstract}
We perform a combined X-ray and strong lensing analysis of RX~J1347.5-1145, one of the most luminous galaxy clusters at X-ray
  wavelengths. {  We show that evidence from strong lensing alone, based on published VLT and new HST data, strongly argues in favor of a complex structure.} The analysis takes into account arc positions, shapes and orientations and is done thoroughly in the image plane. The cluster inner regions are well fitted by a bimodal mass distribution,  with a total projected mass of $M_{tot} = (9.9 \pm 0.3)\times 10^{14} M_\odot/h$ within a radius of $360~\mathrm{kpc}/h$ ($1.5'$). Such a complex structure could be a signature of a recent major merger as further supported by X-ray data. A temperature map of the cluster, based on deep Chandra observations, reveals a hot front located between the first main component and an X-ray emitting South Eastern sub-clump. The map also unveils a filament of cold gas in the innermost regions of the cluster, most probably a cooling wake caused by the
motion of the cD inside the cool core region. A merger scenario in the plane of the sky between two dark matter sub-clumps is consistent with both our lensing and X-ray analyses, and can explain previous discrepancies with mass estimates based on the virial theorem.

\end{abstract}


\begin{keywords}
Galaxies: clusters: general --
                Galaxies: clusters: individual: RX~J1347.5-1145 --
                X-rays: galaxies: clusters --
                Gravitational lensing
\end{keywords}     

\section{Introduction}  
\label{intro}  
  Accurate determination of total mass of galaxy clusters is important
  to understand properties and evolution of these systems, as well as
  for many cosmological applications. Gravitational lensing, through multiple image systems (strong lensing) as
  well as from distortions of background sources (weak lensing), provides a
  reliable method to determine the cluster  mass,   which is  independent of
  the equilibrium properties of the cluster \citep{mel99}. The lensing  mass
  determination can be compared to estimates based on measured X-ray surface
  brightness and temperature, which is instead based on the assumption of
  hydrostatic equilibrium  \citep{sar88} or to dynamical estimates, which rely on the assumption of virialized systems. Combining optical, X-ray and radio observations of galaxy
  clusters is a major tool to investigate their intrinsic properties. In particular, the comparison of lensing and X-ray studies can give fundamental insights on the dynamical state of
  the galaxy clusters (see for example \citet{allen2002, def+al04}), on the
  validity of the equilibrium hypothesis and on their 3-dimensional structure
  \citep{def+al05, ser+al06, ser2007}. 

 RX J1347.5-1145 ($z=0.451$) is one of the most X-ray luminous and massive galaxy cluster known. This cluster has been the subject of numerous  X-ray \citep{schindler95,sch+al97,allen2002,gi+sc04}, optical (Sahu et al.1998) and Sunyaev-Zeldovich (SZE) effect studies \citep{kom+al01,kit+al04}. Formerly believed to be a well relaxed cluster, with a good agreement between weak-lensing \citep{fi+ty97, kli+al05}, strong lensing \citep{sahu98} and X-ray mass 
estimates \citep{sch+al97}, more recent investigations revealed a more 
complex dynamical structure. { In particular}, a region of enhanced emission in the South-Eastern quadrant was first
detected by SZE effect observations \citep{kom+al01} and later confirmed by
X-ray observations that also measured an hotter temperature for the
excess component \citep{allen2002}. This feature  has been interpreted as an indication of a recent 
merger event \citep{allen2002,kit+al04}.  {Furthermore}, a spectroscopic survey on
the cluster members   found  a velocity dispersion of $910 \pm 130\
\mathrm{km~s^{-1}}$, which is  significantly smaller than that derived from
weak lensing, $1500 \pm 160\ \mathrm{km~s^{-1}}$ (Fisher \& Tyson 1997), strong
lensing, roughly 1300 $\mathrm{km~s^{-1}}$ and X-ray analyses $1320 \pm 100\ \mathrm{km~s^{-1}}$ \citep[see their table~4]{co+kn02}.
A major merger in the   plane  of the sky was proposed as a likely scenario to 
reconcile all measurements \citep{co+kn02}. 

In this paper, we further investigate the merger hypothesis  by performing a combined strong lensing and X-ray analysis of archive data. 
We perform a strong lensing investigation based on a family of multiple 
  arc candidates first proposed in 
Brada{\v c} et al. (2005)   using  deep VLT observations.  
Differently from previous studies, we take care of performing 
the statistical analysis in the lens plane, which is a more reliable approach
than the source-plane investigation when working with only one multiple image
system. We further refine our analysis by taking
into account   not only the image positions, but also the shape and
orientation of the arcs.   In addition, we exploit {\it Chandra} observations to gain additional
  insights into the dynamical status of the cluster, through  spectral and
  morphological analyses of the X-ray halo, and to discriminate between different
  evolutionary scenarios. 

The paper is organised as follows. Section~\ref{sec:data} discusses the
strong lensing image candidates selection from archive VLT and {  HST} data.  
Section~\ref{pdsa} describes the statistical method used in the lensing analysis whereas Sec.~\ref{Sec:xray} is devoted to the X-ray data
analysis. Section~\ref{sec:dis} discusses the merger hypothesis. Summary and conclusions are presented in Sect.~\ref{sec:summary}. Throughout this paper we use a flat model of universe with a cosmological constant  with $\Omega_m=0.3$ and $H_0=70\ \mathrm{km\ s}^{-1} \mathrm{Mpc}^{-1}$. This implies a linear
scale of $5.77\ {\rm kpc}/\arcsec$ at the cluster redshift.

\section{Optical data: cluster members and arc candidates}  
\label{sec:data}

\begin{figure}
   \resizebox{\hsize}{!}{\includegraphics{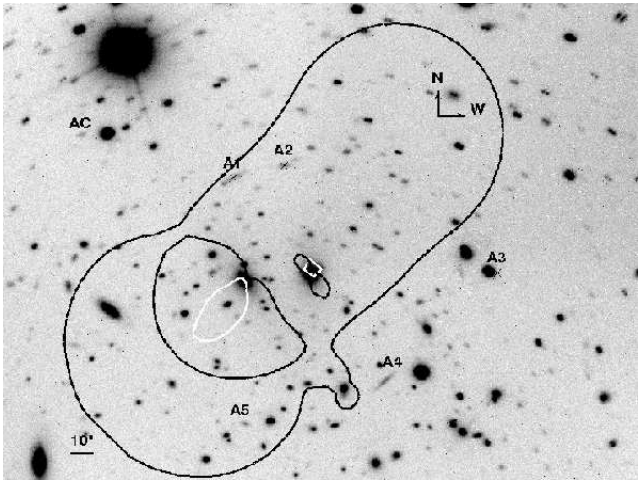}}
   \caption{VLT-FORS1 image of the cluster RX~J1347-1145. {  The black lines are the critical lines for the bimodal
       model (see Sect. 4) corresponding to the source redshift of Arc A4, $z = 1.76$. The white lines are the critical lines
       obtained for Arc A1 ($z = 0.806$).} }
   \label{fig1}
\end{figure}

\begin {figure}  
\resizebox{\hsize}{!}{\includegraphics{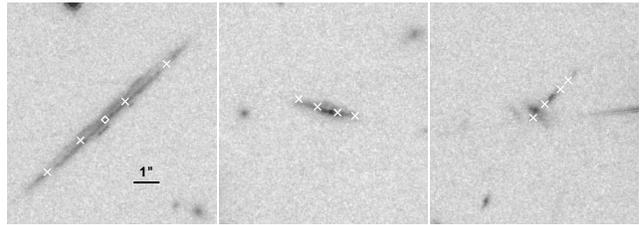}}
\caption{HST images of arcs A4, A5, AC { (in the F814W band)}. On the left, we show the test
    curve A4; in the middle and right panel, we show the reference arc A5 and
    AC, respectively. The diamond corresponds to the observed arc A4
    centroid. The {  crosses} represent the counter-images of the four points
    sampling the test curve A4. }
\label{fig1bis}
\end {figure}

{  The lensing analysis was based on archival VLT data, performed using the same dataset of Brada{\v c} et al. (2005),
 and more recent HST data.
The VLT data were obtained with FORS1} in high resolution mode (pixel scale
0.09\arcsec, total field of view $\sim3.2 \arcmin \times 3.2\arcmin$), using
UBVRI Bessel filters. The seeing in the $I$-filter is of $0.57\arcsec$. Details on the data reduction can be
found in \cite{bradac2005a}. We generated object catalogs for each band, using SExtractor v2.4.4 to measure photometric and geometrical parameters. 
 
  So far, more than  five arc candidates for this cluster have been reported in 
the literature (see Fig.\ref{fig1}). The first two  (A1 and A4) were
discovered by Schindler et al. (1995); later, HST STIS images
revealed three additional ones, i.e. A2, A3, A5  \citep{sahu98}. Recently, Brada{\v c} et al. (2005) reported the discover of several  new arc candidates
(AC, B1 B2, C, D1, D2, D3, D4, E) using VLT data.  
  Arc A1 ($z = 0.806$) is the only one with measured spectroscopic
 redshift (Allen et al. 2002). Despite A3 could appear as a likely counter arc, this
identification is still pretty uncertain.
For the rest of the arcs only photometric redshift estimates are available so
 far \citep{bradac2005a}. Thus, based on the existing literature we decided
 to start to use
the most likely multiple-image system, e.g. the A4-A5 arcs, to model the cluster mass distribution. 
We then searched for additional images.  Following Brada{\v c} et al. (2005), we compared photometric properties and used flux measurements in UBVRI to distinguish different arc families. 
We inspected the galaxies having I-band magnitude up to three magnitudes fainter than the BCG and with a distance to the BCG smaller than $1^{\prime}$. 
We confirm that the most plausible counter arc candidate is AC, giving a system of
three images A4, A5 and AC (see Fig.~\ref{fig1}) which have consistent multi-band colors (using
5" aperture magnitudes) and surface brightness within the errors. Such a system is likely due to a source redshift at $z_{\rm s}\simeq 1.76$ \citep{bradac2005a}. It is important to stress that, while we can not exclude that there are additional multiple-image systems, our
analysis indicates that there are no additional lensed images belonging to the
A4-A5 system within the inspected region and above our detection limit. 

{  New observations from HST, in the F475,
F814 and F850 bands, became recently available (Proposal id. 10492). We analysed
the new data to check whether they confirmed out findings. The three
images, obtained with the ACS camera, are not significantly deeper than
the VLT data, but the higher spatial resolution (0.05" FWHM) allows to
better disentangle nearby sources. The analysis confirms that A4, A5
and AC have consistent colors. In particular AC could now be resolved
in a tangential arc and an overlapping source, see Fig.~\ref{fig1bis}. We find that while the arc is fully compatible with the other images, the overlapping source
is not, thus strengthening the results based on VLT data.}


\section{Statistical analysis}  
\label{pdsa}  

We consider two density profiles to model the mass components: the first one for the cluster-sized halos and the second one for the galaxy-sized objects. We model the main mass (dark matter plus baryonic intra-cluster medium) components as softened power law ellipsoids \citep{keeton2001}. The projected surface mass density at the projected position $\mathbf{x}$ in the plane of the sky is as usual expressed in terms of the convergence $\kappa$, i.e the ratio of the local surface mass density of the lens $\Sigma(\mathbf{x})$ to the critical surface mass density $\Sigcr = c^2 D_s/ (4\pi G D_d 
D_{ds})$, being $D_s$, $D_d$ and $D_{ds}$ the source, lens and the lens-source angular diameter distances, respectively. For a general softened power law ellipsoidal model, 
$$
  \kappa(\xi) \equiv {\Sigma(\xi) \over \Sigcr}={1 \over 2}\,{b^{ 2-\al} \over 
    \left(\theta_c^2+\xi^2\right)^{1-\al/2} }\, \quad 
  \mbox{where}\quad \xi^2 = x_1^2+x_2^2/q^2\ 
\label{nsie} 
$$
where $b$ gives the mass normalisation, $q$ is the projected axis ratio,
$\alpha$ is the slope index, $\theta_c$ is a core radius and $\xi$ is an
elliptical coordinate. We will adopt, for the main dark components, the  non singular isothermal ellipsoid (NSIE) which has $\al$ fixed to 1. 

We model the galaxy-sized  (dark matter plus stellar content) halos as pseudo-Jaffe models \citep{keeton2001}. The pseudo-Jaffe model is obtained by combining two softened isothermal ellipsoids in such a way that the projected density falls as $\xi^{-3}$ outside the cut-off radius $\theta_{cut} (> \theta_c)$.

We performed a $\chi^2$  fit taking care of both the contribution from  the image positions, $\chi_{img}^2$, and the arc shapes and orientations, 
$\chi_{arcs}^2$, so that $\chi^2 = \chi_{img}^2+\chi_{arcs}^2$.  
{  We evaluated them in the {\it image plane} \citep{koch91} using the Gravlens code\footnote{The software is available via the web site:http://cfa-www.harvard.edu/castles}  \citep{keeton2001}. }
The $\chi_{img}^2$ accounts for the agreement between the observed positions of the arc centroids and their predicted values. Computing the $\chi^2_\mathrm{img}$ in the image plane is the most reliable technique when dealing with just one multiple image system. In fact, the
alternative procedure of estimating $\chi^2$ in the {\it source plane} \citep{kayser90}, despite of being
computationally faster, does not take into account how many images make up the
observed system, and might prefer in a flawed way mass models that yield a
good fit adding   fictitious  images. 

We took into account the shape and orientation of the observed arcs sampling
each curve in a number of points \citep{keeton2001}. 
A useful $\chi_{arcs}^2$
takes into account the distances between the fitted points of the
counter-images of a test curve and the data points sampling the other
reference curve. Such an approach is best suited to exploit the A4-A5-AC
system, where the arcs are elongated and well defined.




Our strong lensing statistical analysis was then implemented considering the system A4-A5-AC as a multiple image system. We considered both the centroid
position and shape and orientation of the arcs as input data.  We chose A4, the longest and better defined arc,  as  test curve,   {  leaving  the others} as  reference curves. The positions of the sampling points are showed in Fig.~\ref{fig1bis}.

The total number of constraints ($N_{const}$) in our analysis is {  $18 = 6+8+4$}, being $3 \times 2$
the centroid positions, $4 \times 2$ the additional sampled positions of the
test arc {  and $2 \times 2 $ the additional sampled positions of the Arc A5} . The image centroid of A4 and A5 are used only for $\chi^2_{img}$, so that
$\chi^2_{img}$ and $\chi^2_{arcs}$ are independent. The number of parameters, $N_{par}$, and the number of degrees of freedom, $N_{dof}=N_{const}-N_{par}$,  for the different models are reported in Tab.~\ref{tablechi}.
{  In our fitting analysis, we assume a positional uncertainty of 0.01'.}

\subsection{One component} 

As a first step, we considered the simplest tentative model, i.e a single main
mass component centred in the neighbourhood of the BCG. Although we performed
the whole investigation in the image plane, we first considered also the
$\chi^2_{src}$ minimisation in the source plane. This allows $i)$ to further check
if other additional images of the chosen lens system are formed and $ii)$ to
allow an easy comparison  with previous works.

A model with a single dark matter component provides a good fit only if two further images are present in the A4-A5-AC system. They appear when we check the $\chi^2_{src}$ fitting result on the image plane. These
  two additional images should be lensed in the field of view of the VLT and HST
  observation, $\sim 0.\arcminf 2$ SW of the BCG, with amplification factors of the
  same order of the other three images and should then be easily detected, if present. This model has been then discarded. This first model
is essentially the same as the one used in Brada{\v c} et al. (2005) as a first step in their fitting procedure.

 We then performed the $\chi^2$ minimisation in the image plane letting all the dark component
parameters vary ({  $N_{dof}=10$}).
A single mass component yields
a very poor fit with a very high $\chi^2_{arcs}$.
The mass component is located 0.$\arcminf 13$ East of the BCG.
In Tab.~ \ref{tablechi} you can see the performance and the properties of this model, together with the other models discussed in the following.

A single dark matter component model is therefore not appropriate to describe the matter content in RX~J1347.5-1145, hence revealing a possible more complex configuration.  Even if the details of the results in the following subsections strongly
relies on the position of the counter-arc AC, the information contained in the
shapes and orientations of A4 and A5 alone indicates a somewhat irregular  
mass distribution.

\subsection{Two components} 
\label{sec:2comp}

We tried to improve the fit by adding a second main mass component. We
  first considered the model with the two main matter components being fixed
  at the positions of the two brightest galaxies ({  $N_{dof}=8$}). This model is similar to the
  one proposed by \cite{allen2002}, 
  where the mass components were constrained  by requiring that the overall
  potential was able to produce the northern Arc 1. It is worth noticing that,
  as \cite{allen2002} themselves remarked, their best model could not explain
  A4 and A5 as images of the same source, as instead suggested by later photometric
  observations.

When we consider the A4-A5-AC system as input data, the image positions are  poorly reproduced and  we get a radial A5 with an extremely high $\chi_{arcs}^2$. Even if we move the position of the second mass component from the second brightest cluster galaxy (SCG) towards  the peak of the
  south-eastern X-ray substructure, the fit is still very poor.  

  We then made an additional step further, and let the position of the second dark matter halo free to vary, while leaving the main one fixed on the BCG ({  $N_{dof}=6$}).
A first acceptable result is that a second component at $\sim 1.\arcminf 2$ 
East of the BCG can improve the fit  substantially. The main improvement
obtained with this  model is that the Arc 5 is now tangential, even if
too much elongated and shifted towards North with respect to what observed. Adding  a second component clearly improves our fit, strongly suggesting  a bimodal structure for the cluster.

We further explored the bimodal structure of the cluster by relaxing also the position of the first main mass component ({  $N_{dof}=4$}), see Fig.~\ref{fig3}. The fit obtained is pretty good, $\chi^2_{tot} = 7.13$. The two dark matter components 
have comparable masses and their centres do not coincide with   either  the
BCG or SCG (see Fig.~\ref{fig3}). The first component is located
North-West of the BCG, the second one South-East (but far away from the
secondary X-ray peak).

\subsection{Multiple components}
\label{best}

To test the reliability of the bimodal scenario and  give an insight on some possible degeneracies, we explored some alternatives accounting for other physical effects. We then explored the effect of adding some further components, by considering a model with three main components, keeping the first two dark matter components fixed and adding a third free component, {  $N_{dof}=2$}. We fixed the first component in the BCG (coincident with the first X-ray peak) and the second one in the second X-ray peak leaving the third one free to vary. 
Even if the image positions are well reproduced, this model does not improve substantially the fit to the observations with
respect to the model with two components. {  In fact, the mass clump centred on the
X-ray peak is very small compared to the others and the global result is
very similar to that with one component fixed and the second one free to vary.}  For all models,  when leaving only one component free to vary, while fixing  the other/s one/s, the free component is always located in the eastern part
  of the system, showing a mass of the same order of the other main component,
  while the second frozen component,  located either in the SCG or in the
  second X-ray peak, tends to be much smaller .

\begin{table*}
\begin{tabular}[c]{lccccccc}       
\hline        
\noalign{\smallskip}
Model & $N_{par}$ &$N_{dof}$& $\chi^2_{img}$&$\chi^2_{arc}$&$\chi^2_{tot}$ & AIC & BIC \\       
\noalign{\smallskip}
\hline              
\noalign{\smallskip}
One Comp   & 8 & 10  &244 &2827& 3061.00   & 3077.00 & 3084.12 \\        
\hline
\noalign{\smallskip}
Two Comp:2 fixed &  10&8  & 228&770& 998& 1018&1026.93  \\
Two Comp:1 free 1 fixed&12 & 6   &39.52&89.15&128.67 & 152.67 &163.35    \\  
Three Comp:2 fixed 1 free&16 & 2& 35.36 &72.84 &108.20 & 140.2 &154.44    \\  
Two Comp:2 free&14 & 4  &0.48&6.65 &7.13 & 35.13 &47.59    \\  
Two Comp:2 free +10 galaxies &16& 2&0.13&6.03  &6.16 & 38.16 &52.41    \\  
\hline
\end{tabular}       
\centering       
\caption{Summary of our lensing models. The $\chi^2$ values, the Akaike (AIC) and Bayesian (BIC) information criteria are reported. Note that $N_{par}$ counts 2 free parameters due to the unknown source position during the fitting procedure.}            
\par\noindent  
\label{tablechi}     
\end{table*}

\begin{table*}      
\begin{tabular}[c]{lccccccc}       
\hline        
\noalign{\smallskip}
Cluster-Size & b (arcmin) & $x_1$(arcmin) & $x_2$ (arcmin) & e & PA (deg) & $\theta_c$(arcmin)&$\theta_{cut}$\\       
\noalign{\smallskip}
\hline              
\noalign{\smallskip}
1DM$_{comp}$   & $0.872 \pm 0.03$ &   $0.333\pm 0.04$  & $0.522\pm 0.03$  & $0.08\pm 0.04$ & $-12.4 \pm 0.2$   & $0.521 \pm 0.03$  &--\\        
2DM$_{comp}$   & $0.747 \pm 0.03$ & $ -0.494\pm 0.02$  &$-0.277\pm 0.01$  &$0.05\pm 0.03$ & $ 27.34\pm 0.09$ & $0.40  \pm 0.03$  &--\\     
\noalign{\smallskip}
\hline 
\noalign{\smallskip}
Galaxy-Size & $\sigma$ (km/s) & $ $ &   &  &    &  $\theta_c$(kpc)&   $\theta_{cut} (kpc) $\\  
\noalign{\smallskip}
\hline  
\noalign{\smallskip}
BCG         & $250\pm 100 $    & [0.0]          & [0.0]           &[0.3] &[-0.10] &  [0.5]       & $ 100 \pm 20$ \\ 
SCG         & $212.5\pm 100 $    & [-0.301]          & [-0.033]           &[0.23] &[33] &  [0.425]       & $ 85 \pm 20$ \\ 
\noalign{\smallskip}
\hline
\noalign{\smallskip}
\end{tabular}       
\centering       
\caption{
{  Parameters for the dark matter halos  and of the two brightest galaxies of our best fit model. Values in square brackets are fixed and not minimised. $e$ is the ellipticity; PA the position angle measured North over East. The coordinate system is centred on the BCG galaxy.} }       
\par\noindent  
\label{tab:dmfit}     
\end{table*}

\begin {figure}  
\includegraphics[width=4.2 cm]{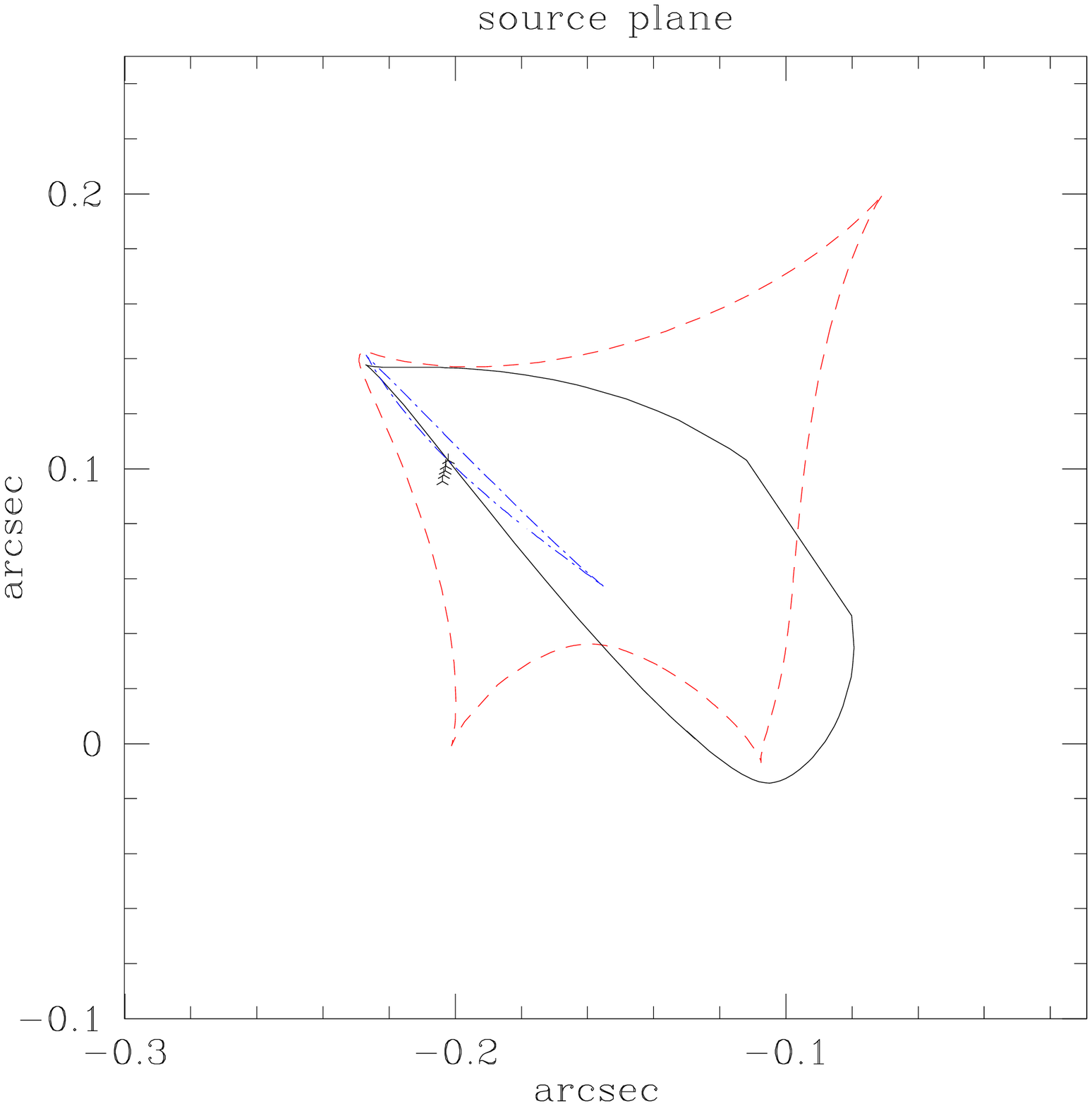}  
\includegraphics[width=4.2 cm]{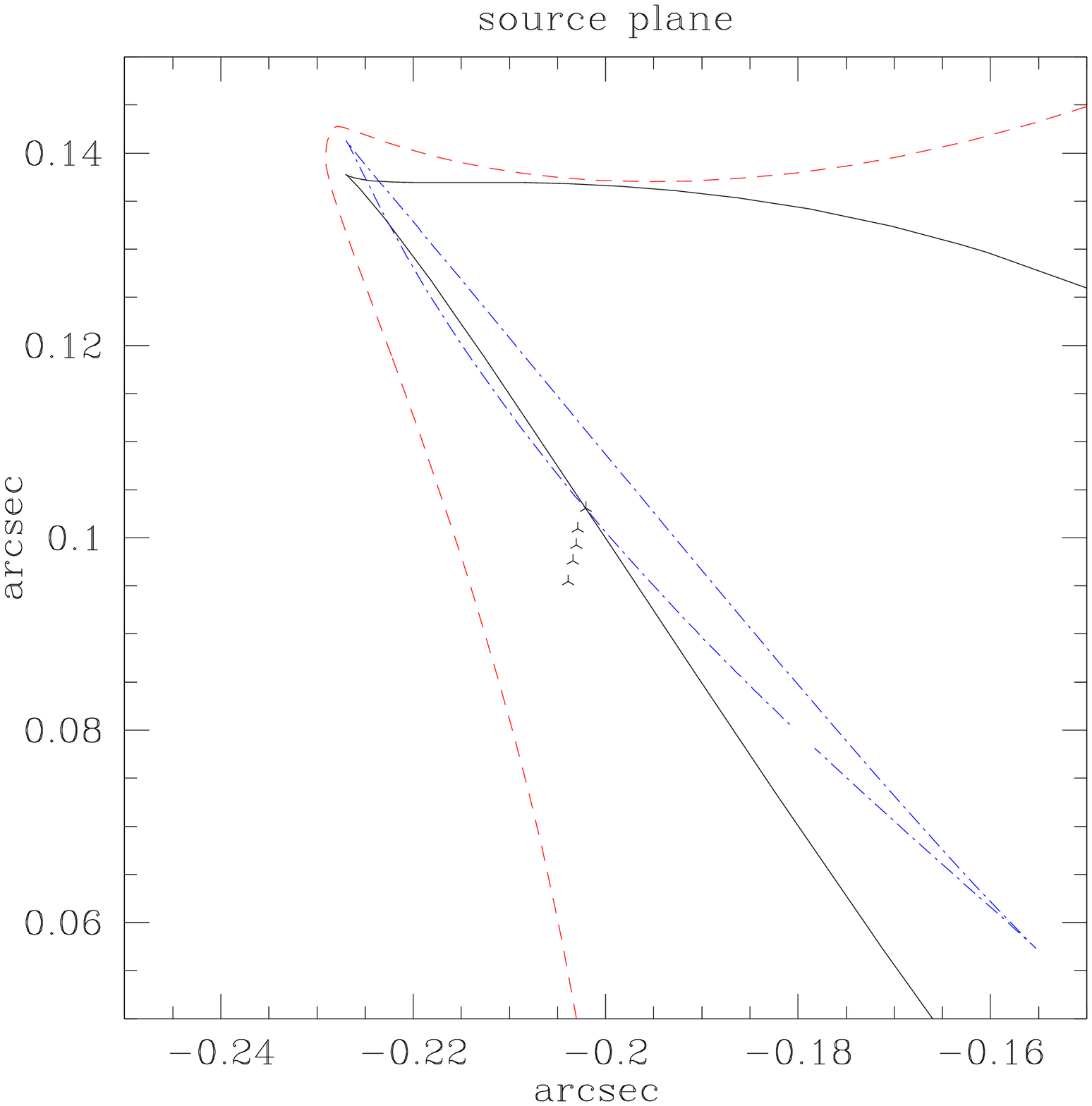}  
 \caption{The position of the source  galaxy { (triangles)} of the arcs A4-A5-AC is shown
   together with the {   tangential caustic (dashed), radial (solid) curves and
   lips (dashed-dotted line)}. The right panel, where the naked cusp {  and the
   source are} clearly visible,
   shows an enlargement of the left panel.}
\label{fig1cau}
\end {figure}  

We eventually considered the effect of galaxy-sized halos. Starting from our bimodal model with two free components, we added 10 galaxy-sized mass components centred at the location of the 10 brightest galaxies of the cluster (selected in the I-band). We expect that only those close to the images can affect the lens configuration. We scale  the p-Jaffe parameters with the luminosity using  the following relations \citep{kne+al96}: $\sigma_0 = \sigma^*(L/L^*)^{1/4}$ and $r_{cut} = r_{cut}^*(L/L^*)^{1/2}$, respectively. We scale the core radius in the same way as $r_{cut}$. 
{  Note that all galaxies, except the BCG and SCG, are assumed to be spherical, see Tab.~\ref{tab:dmfit}.} The proportionality constants $\sigma^*$ and $r^*_{cut}$ are then fitted together with the parameters of the main mass components, so that $N_{dof}=2$.  {  The critical curves for this model are plotted in Fig.~\ref{fig1} (where we also show the critical curves corresponding to the Arc A1 redshift, $z=0.806$) and the caustics in the source plane are shown in Fig.~\ref{fig1cau}. The best fit values are listed in Table~\ref{tab:dmfit} and the surface mass density is plotted in Fig.~\ref{fig3}. The errors on the parameters were estimated performing 100 Monte Carlo simulations. 
} 
Adding  galaxy-sized halos causes a shift in the positions of  the total mass distributions, towards the two main cluster galaxies. We note that when we introduce a galaxy-halo located at the SCG, the peak of the second main component moves South-East. Figure~\ref{fig1bis} reports three enlargements showing the data and fitted points for the system A4-A5-AC.


\subsection{ Model comparison} 
\label{degener}

Lensing information strongly supports a bimodal structure. The reliability of this result appeared by considering different alternative scenarios, which are summarised in Table \ref{tablechi}. To further asses the goodness of our models, we exploited the Akaike (AIC) and the Bayesian information criteria (BIC) (see Liddle 2004 and references therein).
AIC is defined as $AIC = \chi^2 + 2 N_{par}$, while $BIC = \chi^2 + N_{par}\log N_{const}$ where $\chi^2$ is the total $\chi^2$ for the model.
The best model is the model which minimises the AIC or the BIC. By looking at Tab.~\ref{tablechi}, we can see that the one component model is
highly disfavoured, while adding a second mass component improves
the fit, since the BIC value substantially drops down, suggesting a bimodal structure for the cluster. When we let one component position vary, the BIC value becomes even smaller and a substantial improvement is obtained when you let both
component positions be free.  Adding the 10 galaxies does not improve considerably the fit, but it helps to better refine
the features of the cluster model. 
In other words, a bimodal model is naturally favoured by our analysis, which 
definitely rules out the case with just one component. 

The consideration of different lens configurations allowed also an insight into model degeneracies. As can be inferred by the previous analysis, the main degeneracy is connected to the positions of the two cluster-sized halos making up the binary lens. The fit does not change in a dramatic way if the two centres are translated keeping the separation fixed at $\sim 1'$.

Our result on the model configuration does not depend on the estimated source
redshift. In fact, a variation of $z_s$ implies a rescaling of the critical
density, but does not affect the positions and the relative mass ratios of the
components of the cluster. We checked that our result does not
  change in a significant way if we employ a different sampling scheme for the
  curves. {  In the same way, adopting either larger or smaller positional errors does not
    alter the fit values.}

\begin {figure*}  
 \includegraphics[height = 5.5cm,width=5.5cm]{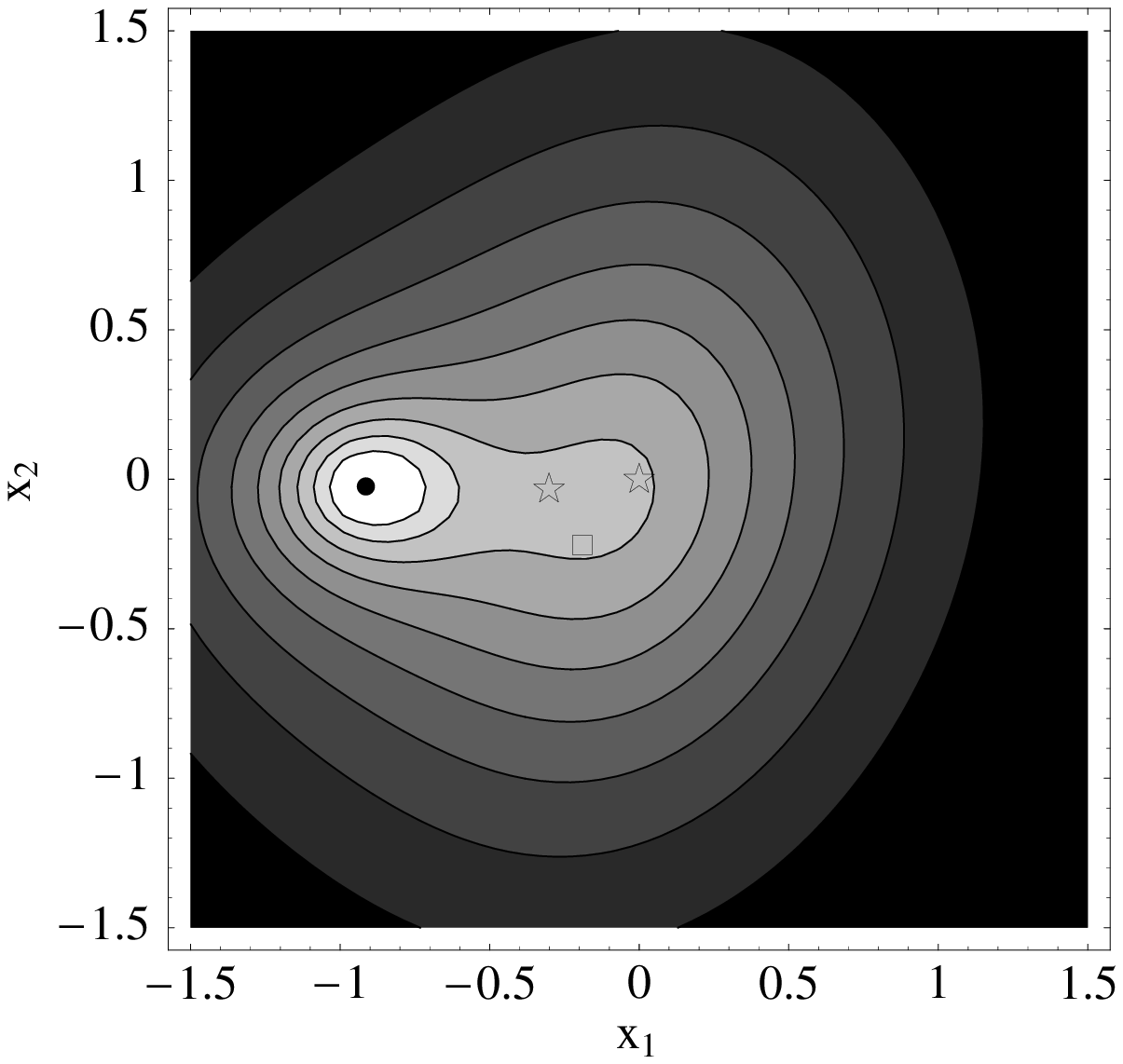}  
 \centering 
  \includegraphics[height = 5.5cm,width=5.5 cm]{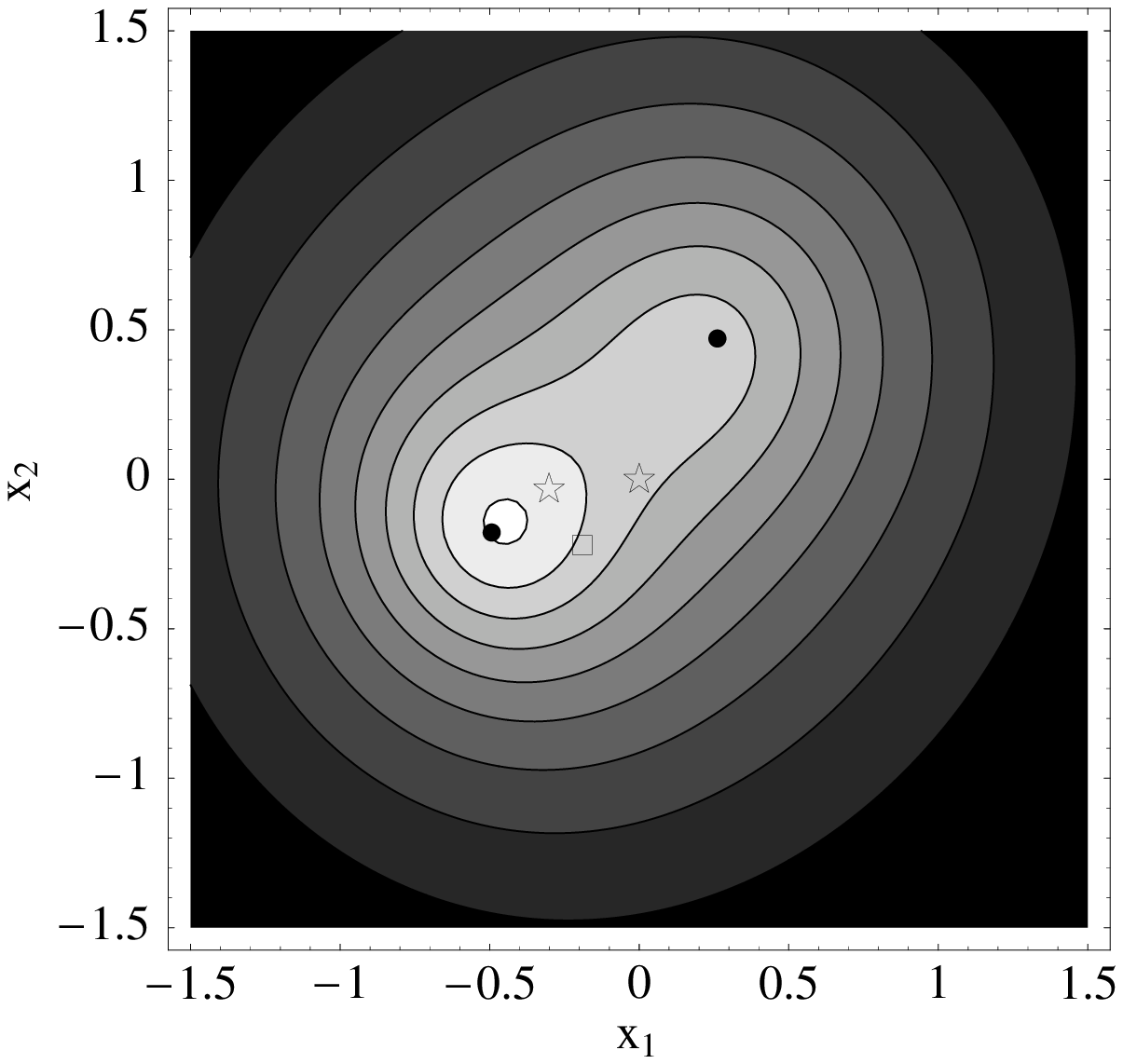}      
  \includegraphics[height =5.5cm,width=5.5 cm]{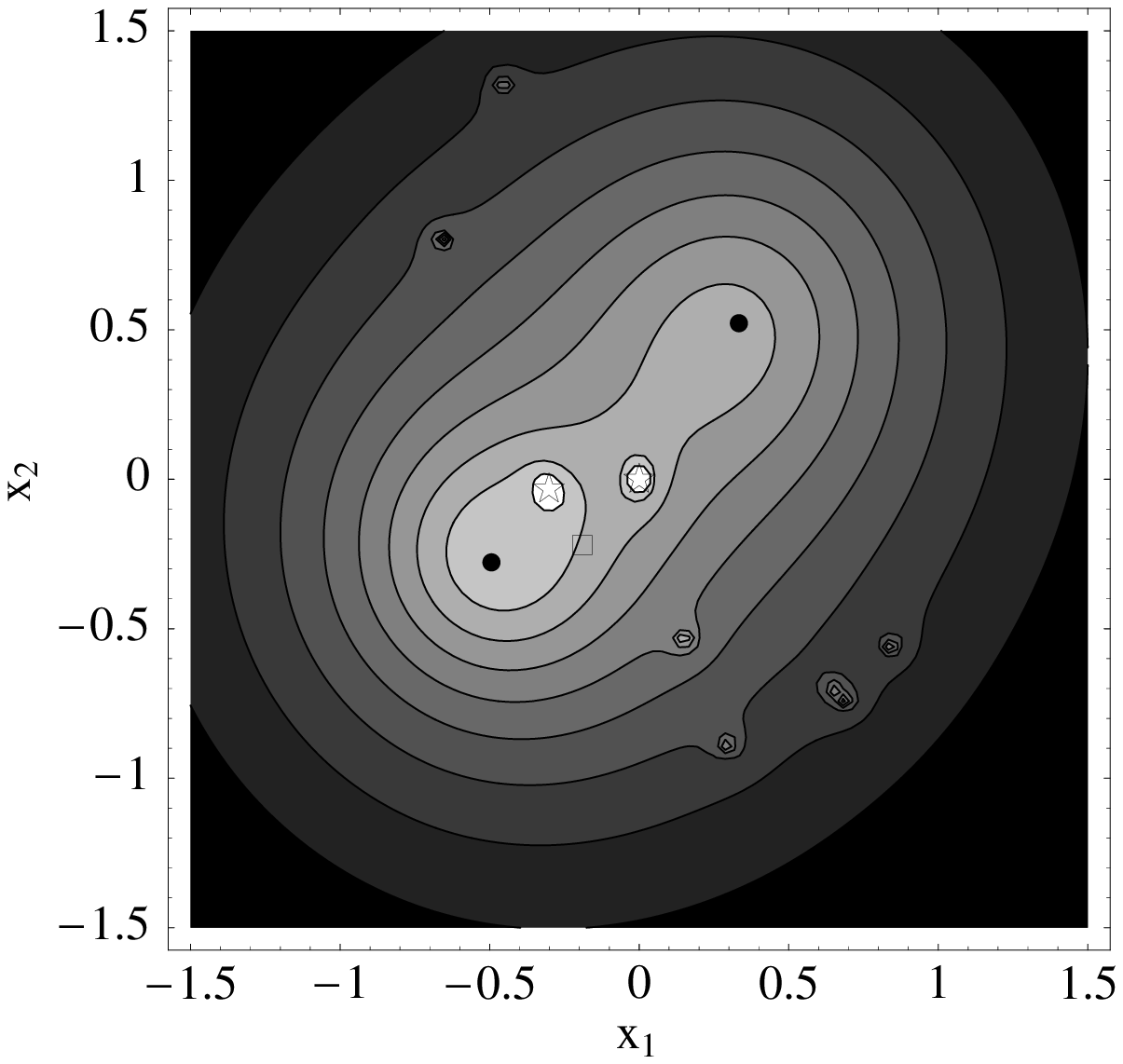}      
\caption  {Projected mass distribution as inferred from the strong lensing
  analysis of the system A4-A5-AC for some   of the  models discussed in
  the text.   Left panel : multiple smooth dark components, one component fixed in the BCG, the second one fixed in the second X-ray peak, and the third one being free to 
  vary.  
  Middle panel: two smooth and free dark matter components. 
  Right panel: our best fit model,   composed by  two smooth dark matter components   plus the  10 brightest galaxies. The coordinate system is centred at the BCG; units are in arcminutes. The surface mass density is in units of $\Sigma_{cr} =
  3100\ h M_\odot\mathrm{pc}^{-2}$; contours represent linearly spaced values
  of the convergence $k$ from 1.3 to 0.5   with a step $\Delta\kappa$ =0.1.
The two star shaped symbols denote the positions of the BCG and the SBG;
  filled dots represent the centres of cluster-sized mass components; the
  position of the SE X-ray sub-clump is given by an empty box.}  
\label{fig3}  
\end {figure*}

\section{X-Ray analysis}
\label{Sec:xray}

\begin{figure*} 
\centering 
\includegraphics[height = 7cm,width=7 cm]{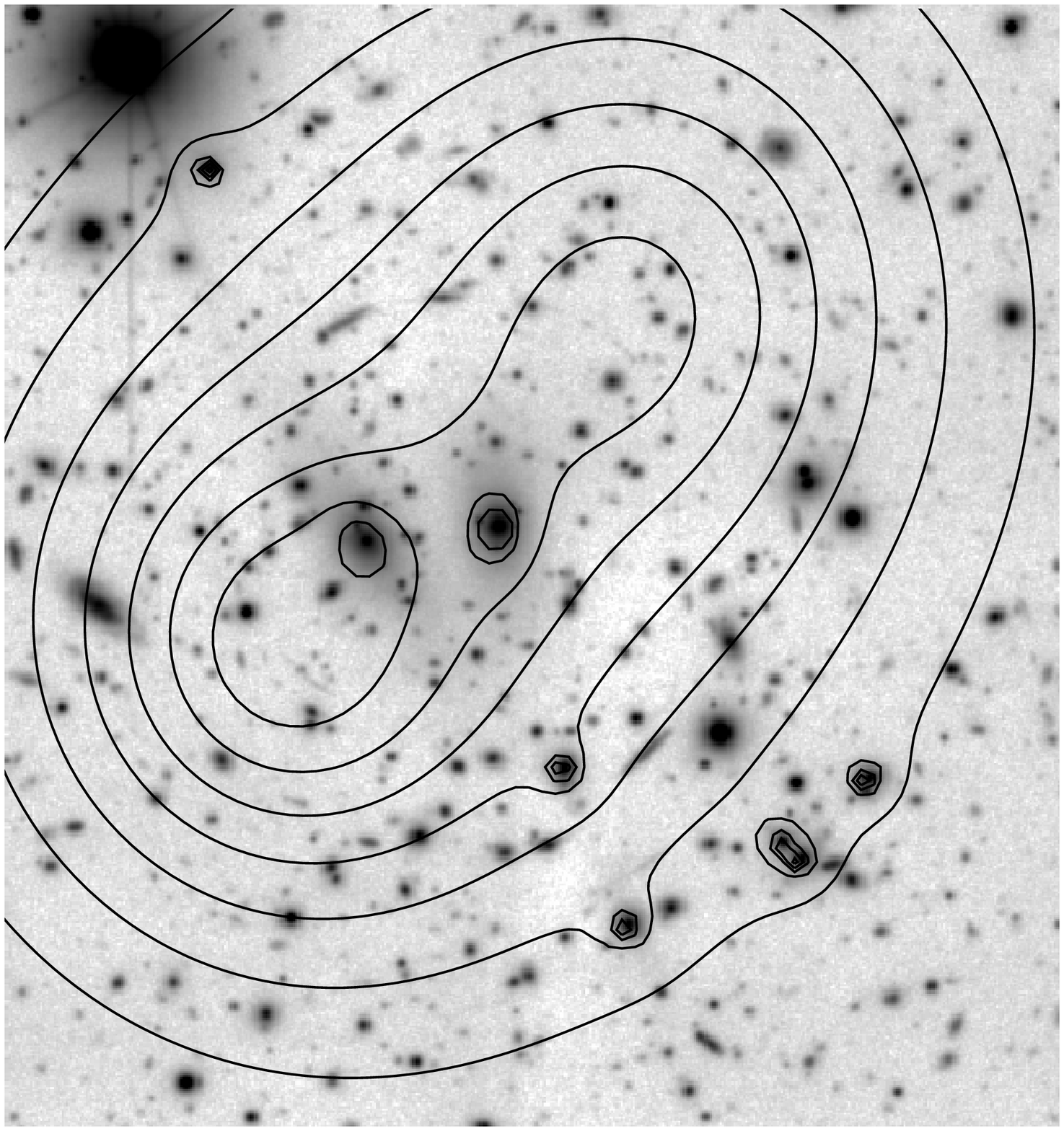}
\includegraphics[height = 7cm,width=7 cm]{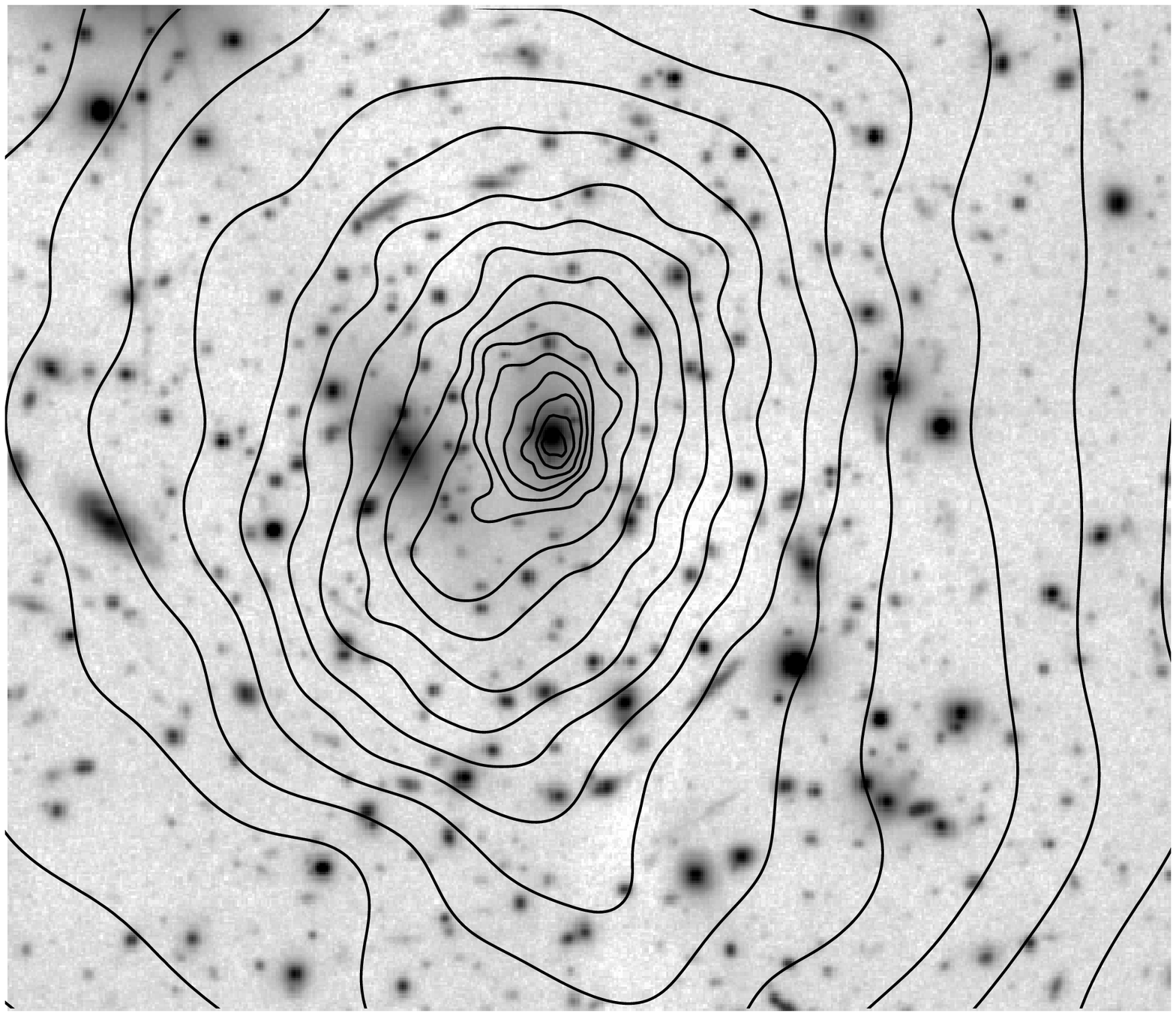}
  \caption{VLT-FORS1 image of RX J1347+1145. Superposed are the contours
    representing the lensing mass distribution (left panel) and the X-ray
    surface brightness (right panel). The units are the same as in Fig.1}
  \label{fig:contours} 
\end {figure*}

We looked for further evidence of the complex structure of RX~J1347.5-1145 by analysing the intra-cluster gas at X-ray wavelengths. RX J1347.5-1145 was first
detected in the X-rays by ROSAT in the All-Sky Survey, which showed and
extremely X-ray bright extended source \citep{schindler95}. Subsequent ROSAT HRI and
ASCA pointed observations revealed it as the most luminous X-ray cluster known at that time, with a bolometric luminosity $L_X\sim 2 \times 10^{46}\ {\rm ergs\ s}^{-1}$ \citep{sch+al97}. 
Based on 
{  a} $\sim 20\ {\rm ksec}$ {\it Chandra} observations, Allen et
al. (2002) reported the discovery of a region of relatively hot, enhanced X-ray emission,
approximately $20\arcsec$ to the south-east of the main X-ray peak (which is located at RA$=13^{\rm h}47^{\rm m}30\fs6$, Dec$=-11\degr45\arcmin09\farcs 3$ (J2000)), 
separated from the main cluster core by a region of reduced emission. 
Later XMM-Newton observations confirmed these results \citep{gi+sc04,Git07}.

In this work we analysed an archival {\it Chandra} observation of
RX~J1347.5-1145, carried out on September, 2003, using ACIS-I {  (already
  published by Allen et al. 2004 within a study of relaxed galaxy clusters}). 
{  The data reduction was performed using CIAO 3.3.0.1.} The net good exposure time, after removing all periods of high background, is of $56\ {\rm ksec}$. This longer {\it Chandra} exposure roughly confirms previous findings detailed above. In the following we will hence exclusively point out some new results, compared to what already published in literature. For the extraction of the azimuthally averaged temperature profile (see \S \ref{sec:X-spec}), 
{  obsID 506 and 507,} with net exposure times of $9\ {\rm and}\ 10\ {\rm ksec}$, respectively, {  were also used}.
In the innermost regions of the cluster the hot gas closely follows the BCG gravitational potential, with the
 X-ray emission approximately oriented in the NS direction (Fig. \ref{fig:contours}). The main X-ray peak is slightly shifted towards SW respect to the centroid of the  overall cluster emission.


\subsection{X-ray Spectroscopy}
\label{sec:X-spec}

\begin{figure} 
\centering 
\includegraphics[width=7 cm]{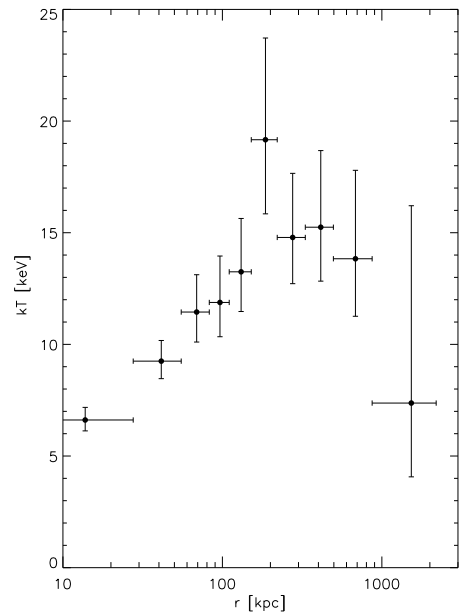}
 \caption{Projected X-ray gas temperature profile in the $0.3-7.0$ keV energy range.} 
 \label{fig:TempProf} 
\end{figure}

\begin{figure*}
\includegraphics[height = 7cm,width=7 cm]{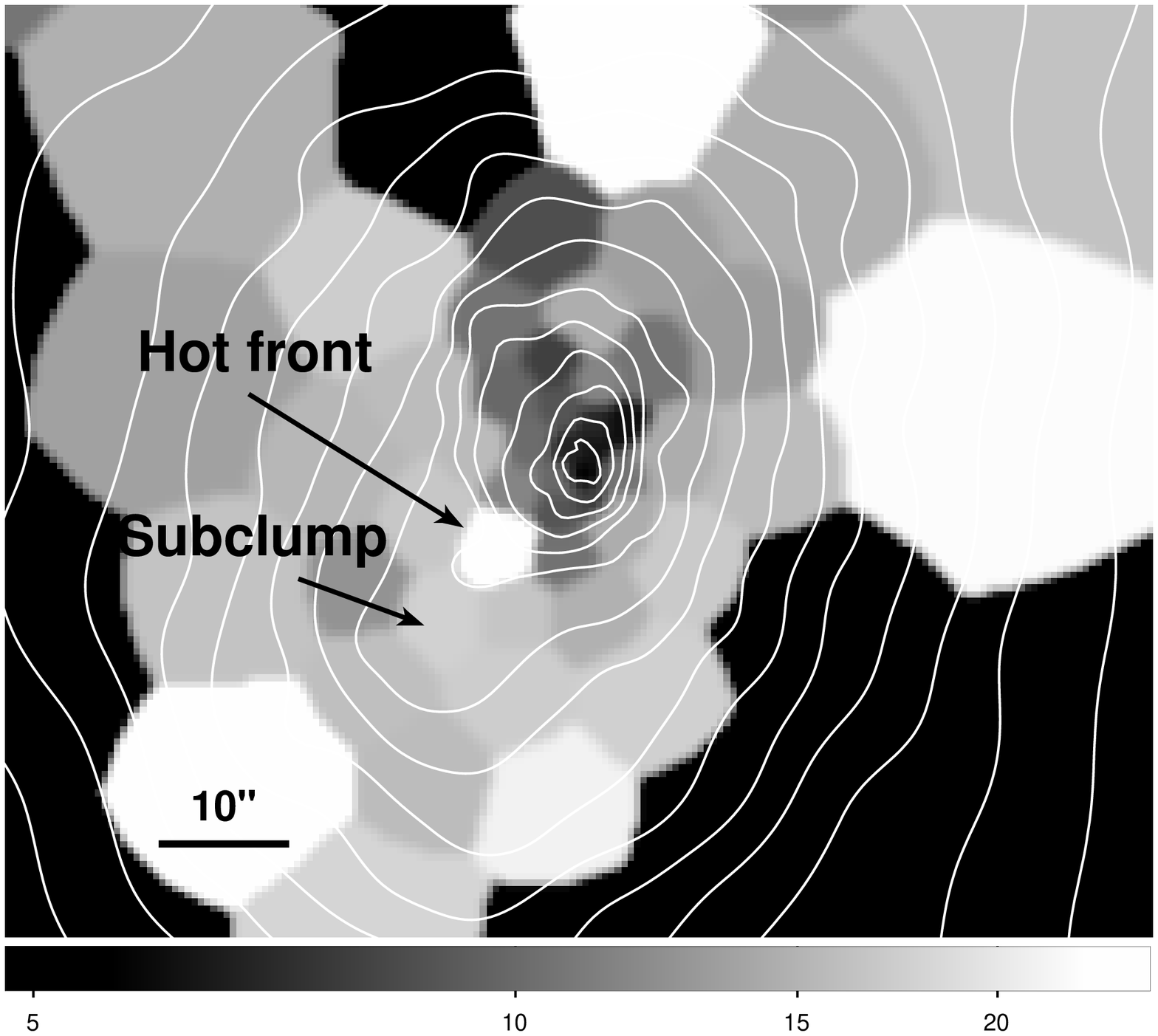}
\includegraphics[height = 7cm,width=7 cm]{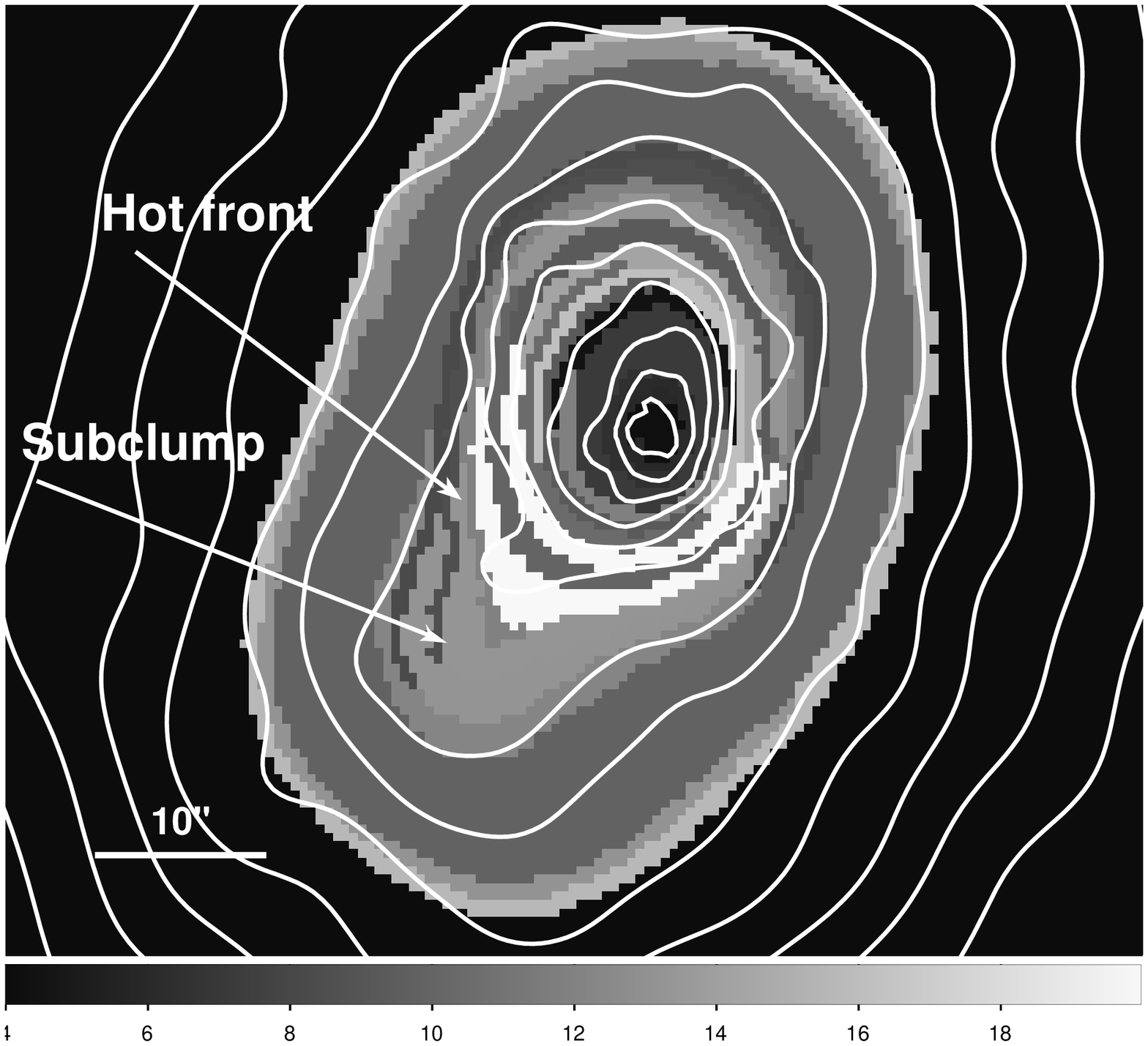}
  \caption{Adaptively binned temperature maps of the central $(1.6\arcmin
    \times 1.6 \arcmin)$ performed using a Voronoi binning algorithm (left
    panel) and a contour binning algorithm (right panel). The temperature scale on
    x-axis is expressed in keV.  Superposed are the logarithmically spaced smoothed X-ray contours.} 
  \label{fig:TempMap} 
\end{figure*}

Throughout this section, spectra have been modelled (in the $0.3-7.0$ keV band) to an absorbed XSPEC isothermal plasma emission code by~\cite{Kaa93}, including the FeL calculations
of~\cite{Lie95} {  (MekaL)}
folded through the appropriate response matrices and
corrected for the ACIS time dependent absorption due to molecular
contamination. The absorbing column density was fixed at the nominal Galactic
value $n_H= 4.85\times10^{20}\ {\rm cm}^{-2}$ \citep{Dic90}. All errors, if not otherwise stated, are $90\%$ 
confidence limits. Within $1.5\arcmin$ from the main X-ray peak (to the full $360 ^\circ$), we
measure a mean emission-weighted temperature of $k T = 11.7\pm 0.5$~keV and a
metal abundance $Z=0.30\pm0.06\ Z_\odot$ 
in agreement with previous Chandra \citep{allen2002} and ASCA \citep{sch+al97}
measurements.

The azimuthally averaged temperature behaviour was inspected by extracting spectra into elliptical annular bins centred at the cluster X-ray main peak, and excluding the South-Eastern quadrant.
We observe a sharp temperature decrease within the central $\sim 200\ {\rm
  kpc}$, consistent with the presence of a massive cool core, and a more
gentle decline in the outer regions~\citep{allen2002,gi+sc04} (see Fig.~\ref{fig:TempProf}). {  The deprojected temperature profile follows the same trend, and does not differ significantly from the projected one shown in the plot.}

{  We then performed} a two-dimensional spectral analysis of the central
$1.6\arcmin\times 1.6 \arcmin$ of the cluster. This was done with the aid of
the WVT binning algorithm by S. Diehl and T.S. Statler~\citep{Die06,Cap03},
and then extracting spectra for each resulting Voronoi cell
(each containing at least 900 photons (S/N$\sim 30$) in the $0.3-7.0\ {\rm keV}$ energy
range, after background subtraction). The resulting temperature map is shown
in the left panel in Fig.~\ref{fig:TempMap}. 

{  We also} produced an additional temperature map using a totally different binning technique: the {\it contour
  binning} algorithm by \cite{San06}. This method chooses regions by following
contours on a smoothed image of the cluster; the generated bins closely follow
the surface brightness distribution. As above, spectra were extracted from each resulting region, each containing at least 900 photons 
The resulting temperature map is shown in the right panel in Fig.~\ref{fig:TempMap}. The relative errors of the temperatures in two resulting temperature maps are of
the order of $10-20\%$, with a slight dependence on the temperature. 
Regions with larger errors were cut off from the final maps.

The two temperature maps reveal a cool core ($kT \sim 5.8\ \mathrm{keV}$) which appears slightly flattened, extending from North to South (see left panel in Fig.~\ref{fig:TempMap}). 
Out to $r\leq 1\arcmin$ the temperature is approximately constant around
$\sim 14.0-17.0\ \mathrm{keV}$. Several much hotter regions are observed, one of which lies surprisingly close to the cool cluster core (the "{\it hot front}" at $9\arcsec$ south-east of the cluster centre).
This {\it hot front} is located in the region of reduced {  X-ray emission that separates the cluster core from the the X-ray sub-clump, which instead shows a slightly lower temperature. The X-ray sub-clump, contrary to what previously thought \citep{allen2002}, does not therefore correspond to the temperature peak in the region. In the {\it hot front} region, neither optical nor X-ray point sources are observed.}
The same {  features are} observed in the temperature map produced with the {\it contour binning} technique
(right panel in Fig.~\ref{fig:TempMap}), showing that 
{  the result cannot be ascribed to} the binning technique.
A trail of cold gas ($\sim 8.0-10.0\ \mathrm{keV}$), approximately $130\
\mathrm{kpc}$ long and extending form the BCG toward N, is also visible. More
than half of the detected photons at that position will be projected from the
surrounding gas; the intrinsic temperature of the cold trail is hence likely
to be lower than the values quoted above. The cold filament is probably a
cooling wake, caused by the motion of the cD galaxies inside the cool core
region~\citep{Dav94,Fab01}.

\subsection{X-ray and SZE face to face}
Our X-ray analysis has revealed hot gas ($kT \stackrel{>}{\sim} 20\ \mathrm{keV}$) in the SE quadrant, associated to a hot front located between the cool core and the SE sub-clump rather than to the sub-clump itself. This front is most probably what remains of a shock front, a clear indication of a recent merger event. \\
SZE effect studies, despite their low spatial resolution, further strengthen the above
conclusions. The SE sub-clump of enhanced emission was first detected through
SZE measurements by Komatsu et al. (2001) . Subsequent   X-ray and SZE  observations revealed a substantial agreement with previous SZE data~\citep[and references therein]{kit+al04}. \\
In order to characterise the high temperature
substructure, Kitayama et al. (2004)  performed a combined X-ray-SZE
analysis. They assumed that the hot substructure was embedded in an ambient
gas identical to that in the other directions, deriving a temperature of
$\sim 28\pm7\ \mathrm{keV}$ for the hot component, in agreement  with our
X-ray measurement for the hot front.

\section{A merger hypothesis}  
\label{sec:dis}  
Our mass estimates for the inner regions in RX J1347+1145 are in substantial 
agreement with previous lensing analyses. From our bootstrap error analysis, we estimated an integrated total projected mass of $(1.42 \pm 0.05)\times 10^{15} 
M_\odot$ inside a radius of $1.5~\mathrm{arcmin}$, consistent with the 
estimate of $(1.2 \pm 0.3) \times 10^{15}M_\odot$ from a combined weak and 
strong lensing analyses in Brada{\v c} et al. (2005). We should point out that our results depend upon the correct redshift determination and identification of members of the multiple image system we use. An error of $\pm 0.5$ on the source redshift implies an error of $\pm 1.0 \times 10^{14}M_\odot $ on the mass estimate. 

A first weak lensing result by Fisher \& Tyson (1997)  gave a total integrated mass of $(1.5 \pm 0.3)\times 10^{15} M_\odot$  within $\sim
5\arcmin$ and a corresponding velocity dispersion of $\sigma_\mathrm{v} =1500 \pm
 160~\mathrm{km~s^{-1}}$ assuming singular isothermal sphere
 model\footnote{The values for $\sigma_\mathrm{v}$ and {\it M} have been converted to our reference $\Lambda$CDM model.}. 
This result was later confirmed considering a larger field of view by Kling et al. (2005), which obtained 
$\sigma_\mathrm{v} =1400^{+130}_{-140}~\mathrm{km~s^{-1}}$   under the same assumptions, while our mass estimate translates into a velocity dispersion $\sigma_\mathrm{v} =1620 \pm 30 ~\mathrm{km~s^{-1}}$. 
Results from both our lensing and X-ray analysis indicate that the cluster is
not in virial equilibrium; the inferred estimate of $\sigma_\mathrm{v}$ has
hence been derived only for comparison with weak lensing results.

Since our analysis showed indications of dynamical complexity, estimates of the cluster mass as inferred from X-ray data under the hypothesis of hydrostatic equilibrium should be treated with caution. Nevertheless, just for comparison with the lensing mass we performed an X-ray analysis excising from the X-ray data the perturbed South-East quadrant{  , using both parametrical (single and double $\beta$ models) and model independent approaches.}
We obtained {  an} X-ray estimate of the mass within
$1.5~\mathrm{arcmin}$ of $(8 \pm 1 \pm 2) \times 10^{14}M_\odot$, where the
first error is statistical and the second one accounts for the uncertainty on
the intrinsic geometry of the cluster, which might be either oblate or
prolate. The total 3-dimensional mass within a spherical region of radius
$1$~Mpc ($2.9~\mathrm{arcmin}$) is $M_{\rm tot}=(1.1\pm 1 \pm 2) \times
10^{15}M_\odot$. These estimates are in good agreement with previous {\it XMM-Newton}
\citep{gi+sc04,Git07} and {\it Chandra} {  \citep{allen2002,All04,All07,Sch07}} X-ray analyses, and are consistent as well with our results from gravitational lensing. However, X-ray mass estimates of perturbed system may be strongly biased. On one hand, neglecting the X-ray South-East sub-clump implies that we are not considering an additional contribution to the total mass, which is consequently underestimated. On the other hand, the temperature in a merging system might be inflated with respect to the temperature of a relaxed cluster, which implies an over-estimate of the mass.

Whereas X-ray and lensing estimates are in good agreement, the dynamical mass estimate is considerably 
smaller \citep{co+kn02}. Cohen \& Kneib (2002) suggested a merger scenario 
to solve this puzzle. Until the merger is complete and galaxy orbits have 
virialized to the new total mass, the dynamical mass estimate would be biased 
toward the mass of the larger clump. Then, a major merger between two clumps 
of comparable mass could reconcile the discrepant observations. The bimodal mass distribution inferred in  our analysis is in agreement with this merging scenario . 
Moreover the merger is possibly happening in the plane of the sky, since the 
redshift distribution of the cluster members does not show any feature along
the line of sight.

The merger scenario is also supported by the X-ray and SZE observations 
of the intra-cluster medium. Both the gas sub-clump and the hot front in the SE 
quadrant are indications of a merger process. The overall agreement between 
the X-ray and lensing mass estimates might suggest that the emitting gas has
  had time to virialize.  
Hydrostatic equilibrium is also supported  by the 
fact that the overall mass, temperature and luminosity of RX J1347+1145 
  relate as expected by a  typical for massive cluster \citep{allen2002}. 
A major merger scenario was also considered in Kitayama et al.(2004), which discussed as an head-on collision of two $\sim 
5\times 10^{14}M_\odot$ clusters with relative velocity of $\sim 
4000~\mathrm{km~s^{-1}}$ would result in a bounce-shock with $k T 
\stackrel{>}{\sim} 25~\mathrm{keV}$, as detected in the SE quadrant.

\section{Summary and conclusions} 
\label{sec:summary}

We  have analysed both the lensing and X-ray properties of 
RX~J1347.5-11.45, one of the most luminous and 
massive X-ray cluster known. Based on 
the analysis of an arc family, photometrically selected, we have estimated
the total cluster  mass distribution, within a radius of $\sim 500\ \rm{kpc}$
from the cluster centre. We performed a $\chi^2$ analysis in the lens plane and scrupulously modelled the 
arc configuration. A model with two smooth dark matter components of similar 
mass   accurately  reproduces the observations  and yields a mass estimate in agreement 
with previous strong and weak lensing and X-ray studies. Our strong lensing analysis suggests a major merger between two sub-clumps of similar mass 
located within the central $300~\mathrm{kpc}$. 

X-ray observations further strengthen our view of a complex structure of the inner regions, revealing a hot front in the 
South-Eastern area, most probably a remnant of the occurred merger.
This merging framework, which arises naturally from our strong lensing model, can also reconcile the observed discrepancy between dynamical mass estimates and X-ray, lensing and
  SZE  ones, which instead give consistent results. Agreement between X-ray 
  and lensing mass estimates further indicates that the gas might have had time to
  virialize. Spectroscopic measurements additionally suggest that the merger
  is taking place in the plane of the sky.  

Whereas the presence of a merger is confirmed on several grounds, its properties are 
though still unclear.  The detailed features of our model are strictly related to the selection of the members of the multiple image system, in particular to the inclusion of the arc AC.  Despite  all candidates should be confirmed spectroscopically,  a photometric analysis suggests that A4 and A5 almost undoubtedly
belong to the same system. Remarkably, based on the shapes and orientations of A4 and A5
alone, without any constraint on AC, we can exclude the presence of a single mass component.
A  spectroscopic confirmation of further arc candidates is the required step  to confirm and accurately define   the
merging  scenario  and to provide a final, more accurate description of the
dynamical state of RX J1347.5-1145.

\section*{Acknowledgments}
We thank Thomas Erben and Sabine Schindler for providing the 
  VLT/FORS data. We thank Charles Keeton for making available the GravLens software and some useful advices.
 MM and MS are supported by the Swiss National Science Foundation. MS also 
acknowledges the financial support of the Tomalla foundation. EDF acknowledges
  also financial support of the VO-Tech project.

\bibliographystyle{mn2e}

\end{document}